\documentclass [prl,reprint,notitlepage,superscriptaddress,floatfix] {revtex4-1}
\usepackage{amsmath}
\usepackage{graphicx}
\usepackage{lmodern}
\usepackage{amsmath}

\usepackage{color}
\usepackage{amssymb}
\newcommand{\beq} {\begin{equation}}
\newcommand{\eeq} {\end{equation}}
\newcommand{\bea} {\begin{eqnarray}}
\newcommand{\eea} {\end{eqnarray}}
\newcommand{\be} {\begin{equation}}
\newcommand{\ee} {\end{equation}}
\renewcommand{\(}{\left(}
\renewcommand{\)}{\right)}
\renewcommand{\[}{\left[}
\renewcommand{\]}{\right]}

\DeclareMathOperator{\sgn}{sgn}

\begin{document}

\title{Reply to ``A strong coupling critique of spin fluctuation driven charge order in underdoped cuprates"}
\author{Yuxuan Wang}
\affiliation{Department of Physics, University of Wisconsin, Madison, WI 53706, USA}
\author{Andrey Chubukov}
\affiliation{William I. Fine Theoretical Physics Institute,
and School of Physics and Astronomy,
University of Minnesota, Minneapolis, MN 55455, USA}
\date{\today}

\begin{abstract}
We reply to the criticism from the authors of arXiv:1502.02782 of the spin-fluctuation scenario for charge order in the cuprates.
The authors of of arXiv:1502.02782 argued that spin-fluctuation exchange cannot give rise to charge order with observed momentum $(Q,0)/(0,Q)$ due to
 the absence of nesting for a half of fermions involved. We explicitly show the instability towards charge order exists even in
 the ``worst-case" scenario of anti-nesting for a half of fermions.
\end{abstract}
\maketitle

In the recent preprint~\cite{mike} the authors  performed the numerical study of superconducting and charge-density-wave instabilities
 in the model of fermions interacting via magnetic fluctuations. They argued that  there is no instability in the charge-density-wave
  channel with momentum $(0,Q)/(0,Q)$ (CDW-x in their notation) because there is no analog of Cooper logarithm  due to the lack of Fermi surface (FS) nesting
    for one half of the pairs of hot spots.  This result contradicts earlier works~\cite{laplaca,charge,debanjan}, including our earlier analytical study~\cite{charge}.
   In this reply, we consider the same model as the authors of~\cite{mike} and, like them, calculate and compare the eigenvalues in the superconducting (SC)
     and CDW-x channels, $\lambda_{sc} (T)$ and $\lambda_{cdw}(T)$ (defined such that the system develops an instability once $\lambda (T)$ becomes greater than one).   To make our point, we 
      only
      consider the ``worst-case" scenario of perfect anti-nesting for one pair of hot spots.
    We show explicitly that $\lambda_{cdw}(T) $ diverges logarithmically at $T \to 0$, in the same way as $\lambda_{sc}$, and
    the ratio 
    $\lambda_{cdw}/\lambda_{sc}$  is just a constant (smaller than one), 
     whose magnitude is determined by the ratio of spin-fermion coupling and Landau damping.
     As the consequence, $\lambda_{cdw}$ crosses $\lambda_{cdw} =1$ at some finite $T_{cdw}$.
    We believe that Ref. \cite{mike} missed this instability because the strength of the Landau damping was taken to be much larger than one would get by calculating it self-consistently.

    We also comment on the 
    validity of neglecting the fermionic self-energy, as it was done in Ref. \cite{mike}. We argue that keeping self-energy is essential even in the non-quantum-critical, Fermi liquid regime for proper treatment of
    the effects due to FS curvature.
     We also comment on the claim in \cite{mike} that other potential difference between their and earlier works is that previous analysis  relied on an expansion around the hot spots, while the
  calculations in Ref. \cite{mike} were performed by integrating over the full Brillouin zone.
  We argue that for the 
   parametrs 
   used in \cite{mike}, the contribution to the momentum integral comes predominantly from the vicinity of hot spots, and thus the expansion around hot spots is a good approximation.

To make explicit comparison with Ref.\ \cite{mike}, we consider 
   the same model as they did, namely fermions 
    interacting by exchanging   magnetic fluctuations peaked at $\bf K =(\pi,\pi)$, and, like them, we neglect fermionic self-energy.
 The  dispersion considered in Ref. \cite{mike} is reproduced in Fig.\ 1.  For this dispersion, the velocities at hot spots separated along $X$ or $Y$ direction are either almost anti-parallel (for one half of hot spots) or almost parallel (for the other half).  For SC channel, the velocities of relevant fermions are always anti-parallel. 
  The form of spin-mediated dynamical interaction used in Ref. \cite{mike} is the same 
  as in earlier works~\cite{acs,charge}:
\begin{align}
&\Gamma_{\alpha\gamma,\beta\delta}({\bf q},\Omega_m)= {\bar g} \vec \sigma_{\alpha\beta}\cdot \vec\sigma_{\gamma\delta}~ \chi ({\bf q} - {\bf K}, \Omega) \nonumber\\
& \chi (k, \Omega) = \frac{1}{{\bf k}^2+\gamma|\Omega|+\xi^{-2}},
\end{align}
where $g$ is the effective spin-fermion coupling constant, $\gamma$ is the Landau damping coefficient, and $\xi$ is the magnetic correlation length.

 To simplify the presentation, we take velocities at hot spots to be either strictly parallel or antiparallel, neglect FS curvature at hot spots,
 and compare the eigenvalues in SC and CDW-x channels by analyzing ladder equations for the corresponding vertices (see Fig.\ 2).
  The SU(2) symmetry of the hot spot model (Refs.\cite{ms,efetov}) implies that the kernel in the
  SC channel and the one in CDW-x channel for anti-parallel velocities are identical.  The difference between two channels is in that for CDW-x channel
  the interaction peaked at $(\pi,\pi)$  takes fermions from the region where Fermi velocities are anti-parallel (region 1-2 in Fig.\ 1)
   to the region where Fermi velocities are parallel (region 3-4 in Fig.\ 1). As the consequence, one needs a composite second-order process to bring fermions back
   into the region where Fermi velocities are antiparallel, see Fig.\ 2c 
   (a similar reasoning has been recently applied to the analysis of the optical conductivity in the cuprates~\cite{optical}).

   The integration over two Green's functions with antiparallel velocities gives rise to a conventional Cooper logarithm $\log ({\omega_{sf}/T})$, 
    where $\omega_{sf} \sim \xi^{-2}/\gamma$, 
    and the
   logarithm comes from 
   the smallest
    momenta and frequencies. To logarithmic accuracy, the eigenvalue/eigenfunction
   equations for superconducting and CDW condensates, $\Phi_{sc} (k_x)$ and $\Phi_{cdw} (k_x)$ are (small $k_x$ is along the FS near hot spots 1 and 2)
\begin{widetext}
   \bea
   &\lambda_{sc} (T) \Phi_{sc} (k_x) = 3 \frac{{\bar g}}{4\pi^2 v_F}\log({\omega_{sf}/T}) \int dk'_x \Phi_{sc} (k'_x) \chi (k_x-k'_x,0) \nonumber\\
   &\lambda_{cdw} (T) \Phi_{cdw} (k_x) = 3\frac{{\bar g}}{4\pi^2 v_F} \log({\omega_{sf}/T}) \int dk'_x \Phi_{cdw}  (k'_x) \chi_{com} (k_x,k'_x,0)
   \label{ch_a}
   \eea
   \end{widetext}
   where  the numerical prefactors are the results of summation over spin indices 
    and the shift by ${\bf K}$ is absorbed into the shift of momenta.
   The interaction is attractive in both SC and CDW channels if we additionally assume that both $\Phi_{sc} (k)$ and $\Phi_{cdw} (k)$
 change sign under ${\bf k} \to {\bf k} + {\bf K}$ (Refs. ~\cite{laplaca,charge,debanjan}).
   
   The composite $\chi_{com} (k_x,k'_x,0)$ is the convolution of two dynamical spin susceptibilities
   and two Green functions of fermions  with parallel velocities:
\bea
&&\chi_{com} (k_x,k'_x,0) = \label{ch_2} \\
&&-\frac{3\bar g}{8\pi^3}\int \frac{d\Omega_m dp_x dp_y}{(i\Omega_m-v_Fp_x)^2}  \chi (k_x-p_x,p_y, \Omega) \chi (k'_x-p_x, p_y, \Omega). \nonumber
\eea
 where momenta $p_x, p_y$ and frequency $\Omega$ are for fermions in region 3-4 where velocities of the two hot fermions are parallel.
 If we approximated $\chi ({\bf q}, \Omega)$ by its static part, or took it on the FS
 (i.e., set $p_x =0$ in the susceptibilities), the integral would
 vanish because of double pole in $1/(i\Omega_m-v_Fp_x)^2$. This is what the authors of Ref. \cite{mike} probably meant when they argued that there is
   no CDW-x instability due to anti-nesting for a half of hot spots.   However, if we use the full forms of $\chi ({\bf q}, \Omega)$, we immediately obtain that the
    integral in (\ref{ch_2}) does not vanish and yields~\cite{note}
 \begin{align}
\chi_{com} (k_x,k_x',0) = A \chi (k_x-k_x', 0)  f(k_x\xi, k_x'\xi), ~A = \frac{3 {\bar g}}{2\pi^2 \gamma v^2_F} \nonumber
\end{align}
 where $f(0,0) =1$ and to a good numerical accuracy, $f(x,y)$ remains close to one for relevant $ x,y = O(1)$.   As a result, $\chi_{com} (k_x,k'_x,0)$ and
 $\chi (k_x-k'_x,0)$ differ just by a constant $A$. Using (\ref{ch_a}), we then find that
  the ratio
 $\lambda_{cdw}/\lambda_{sc} = A$ is independent on $\xi$.
  This result implies that $\lambda_{cdw}$  has the same logarithmic dependence on $T$ as $\lambda_{sc}$
  and 
   therefore must cross one
    at some finite $T_{cdw}$. This temperature is numerically, but not parametrically smaller than superconducting $T_c$.

   The magnitude of $A$ depends on the interplay between
 spin-fermion coupling and Landau damping coefficient.  In  self-consistent calculations within the spin-fermion model $\gamma$ by itself is expressed
   via ${\bar g}$ as $\gamma=4\bar g/(\pi v_F^2)$. (Refs.~\cite{acs,ms}), hence $A = 3/(8\pi)$ is just a number. The value of $A$ increases if one additionally assumes that
    CDW-x order emerges from some pre-existing pseudogap state which reduces Landau damping~\cite{fl*,atkinson,tigran} due to a reduction of a
     low-energy fermionic spectral weight in the hot regions.
      The Landau damping coefficient used in Ref.~\cite{mike} was not obtained self-consistently and was taken to be $20 \rm eV^{-1}$ for
      ${\bar g} = 0.9\rm eV$ and $v_F = 0.78\rm eV$, which is about ten times larger than self-consistent value.
        This yields a much smaller $A = 0.01$, which is probably the reason why CDW-x instability has not been observed in Ref. \cite{mike}.

\begin{figure}
\includegraphics[width=0.78\columnwidth]{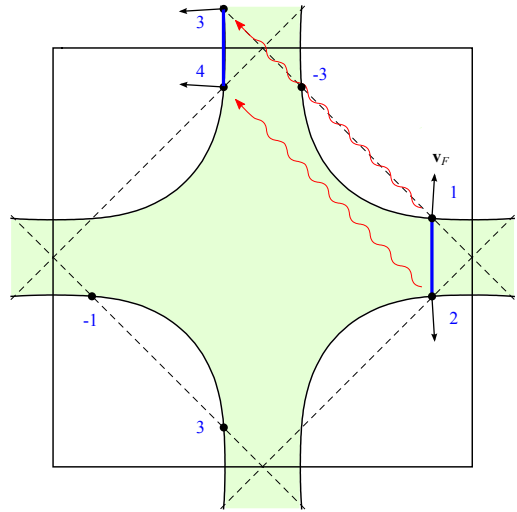}
\caption{The Brillouin zone, the magnetic Brillouin zone, and FS using the same dispersion as in Ref.\ \onlinecite{mike}. Hot spots 1,2 and 3,4 are defined as points on the Fermi surface that intersects with the magnetic Brillouin zone. The red wavy lines represents the interaction mediated by spin fluctuations of momentum $(\pi,\pi)$.}
\end{figure}
 \begin{figure}
\includegraphics[width=.86\columnwidth]{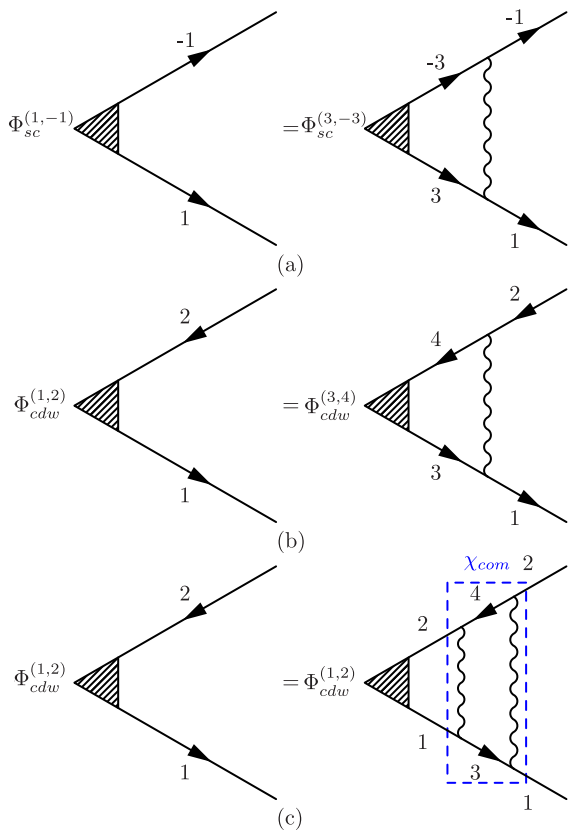}
\caption{The superconducting vertex $\Phi_{sc}$ [Panel (a)] and the CDW-x vertex $\Phi_{cdw}$ [Panels (b) and (c)] in the ladder approximation. Panel (b) shows the relation between CDW condensate formed by fermions at hot spots 1 and 2 and the one formed by fermions at hot spots 3 and 4. In Panel (c) we integrated out fermions near hot spots 3 and 4 and obtained the self-consistent equation for the CDW condensate formed by fermions at hot spots 1 and 2.}
\label{vs}
\end{figure}

We next 
 discuss the effects of fermionic self-energy and FS curvature.  The self-energy was neglected in Ref. \cite{mike}
(and in this respect their approach is not a truly strong coupling).  
It
 plays 
 the central
  role
 in relating SC and CDW instabilities in the quantum-critical, non-FL regime when relevant $T > \omega_{sf}$.  This regime requires a separate consideration~\cite{charge} and we will not discuss it here.   At  energies below $\omega_{sf}$, which we consider here, the self-energy
  has a Fermi liquid form and gives rise to mass renormalization $m^*/m = 1 + \lambda$, where $\lambda = 3 {\bar g} \xi/(4 \pi v_F)$.  
  The self-energy is not overly important for Eq.  (\ref{ch_a}) as it only 
   adds $1/(1+\lambda)$ to the overall factor in the r.h.s.  but does not affect the logarithm.  The inclusion of the self-energy 
    is, however, essential for
     the analysis of the distractive effect of FS curvature on $T_{cdw}$.  One can easily check that the curvature (the term $\kappa k^2_x$ in the fermionic dispersion near hot spots 1 and 2), cuts the Cooper logarithm in the CDW channel at a frequency of order $\kappa \xi^{-2}/(1 + \lambda)$. Using $\kappa \sim v^2_F/E_F$, we find that this frequency is of order $\omega_{sf} ({\bar g}/E_F)/(1 + \lambda)$.  The logarithmical dependence on $T$ in the CDW channel survives as long as
 $ ({\bar g}/E_F)/(1 + \lambda) \ll 1$.  Without 
 $1+\lambda$ term coming from self-energy,
   this condition is
   satisfied only 
   if ${\bar g}\ll E_F$, while for parameters in Ref. \cite{mike} ${\bar g} \sim E_F$.  As a result, without self-energy,
    the FS curvature strongly reduces $T_{cdw}$. 
     With self-energy, the $\log \omega_{sf}/T$ behavior in CDW-x channel 
      definitely 
      holds
      at large enough $\xi$, when $\lambda \gg1$. As the consequence, at large enough $\xi$,
      $T_{cdw}$ is only weakly affected by the 
        FS curvature.
         As $\xi$ decreases, the effect of the curvature gets stronger, $T_{cdw}$ decreases and eventually vanishes at some finite $\xi$, as we argued before~\cite{charge}
  The same behavior due to FS curvature holds for charge bond order with momentum $(Q,Q)$ (Refs.\ \cite{efetov, charge}).

Finally, we discuss the validity of the expansion around hot spots. From Eq.\ (4) we see that the relevant momenta $k_x, k_x'$ (which, we remind, are 
deviation from  hot spots 1, 2 
along the FS) are of order $\xi^{-1}$. 
The typical momenta $p_x$ and $p_y$  in  Eq.\ (3)
are also of order $\xi^{-1}$.
   Taking, e.g., $\xi = 3a$, we obtain~\cite{comm} that 
   typical momentum deviation from a hot spot is $\sim 0.06\times 2\pi/a$. This is a fairly small momentum range.
     As a comparison, for the dispersion taken in \cite{mike}, the separation between neighboring hot spots 
     is much higher: $0.2\times 2\pi/a$. In this sense already for $\xi =3a$, SC and CDW instabilities come from vicinity of hot spots,
      and this is even more so for $\xi_{\rm AF}=5a, 10a$ also
      considered in \cite{mike}. 

To summarize, we have shown explicitly that 
 anti-nesting for a half of hot spots does not prevent the instability towards CDW-x order (the one with momentum $(Q,0)$ or $(0,Q)$) 
  as the corresponding eigenvalue $\lambda_{cdw}$ still diverges logarithmically at low $T$.  We believe that such an instability was 
   not found in the numerical analysis in Ref.\ \cite{mike} because (i) these authors used a much larger value of Landau damping than the one obtained in self-consistent calculations, and (ii) they  neglected fermionic self-energy and
    as a result overestimated the reduction of $T_{cdw}$ by FS curvature. 

We thank  M. Norman
 for fruitful discussions. The work was supported by the DOE grant DE-FG02-ER46900.


\begin{thebibliography}{99}
\bibitem{mike}V. Mishra and M. R. Norman, arXiv:1502.02782 (2015).
\bibitem{laplaca} S. Sachdev and R. La Placa, Phys. Rev. Lett. {\bf 111} 027202 (2013);
 A. Allais, J. Bauer and S. Sachdev, Phys. Rev. B {\bf 90} 155114 (2014).
\bibitem{debanjan}  D. Chowdhury and S. Sachdev, Phys. Rev. B {\bf 90}, 134516 (2014).
\bibitem{charge} Y. Wang and A. Chubukov, Phys. Rev. B {\bf 90} 035149 (2014).
\bibitem{acs} Ar. Abanov, A. V. Chubukov, and J. Schmalian, Adv. Phys. {\bf 52}, 119 (2003).
\bibitem{ms} M. Metlitski and S. Sachdev, Phys. Rev. B {\bf 82}, 075128 (2010).
\bibitem{efetov} K. B. Efetov, H. Meier and C. Pepin, Nat. Phys. {\bf 9} 442 (2013); H. Meier, C. Pepin, M. Einenkel and K.B. Efetov, Phys. Rev. B {\bf 89}, 195115 (2014).
\bibitem{optical} S. A. Hartnoll, D. M. Hofman, M. A. Metlitski, and S. Sachdev,
Phys. Rev. B 84, 125115 (2011);  A. V. Chubukov, D. L. Maslov, and V. I. Yudson,
Phys. Rev. B 89, 155126 (2014).
\bibitem{note} The evaluation of this integral requires some care. The double pole can be regularized either by adding a small external frequency $\delta\omega\to0$, or by adding a small deviation $\delta q\to 0$ from ${\bf Q}$ to the total momentum, and the two limits do not commute. Since we are only interested in static order parameters, we take $\delta\omega\to0$ limit first.
\bibitem{fl*} D. Chowdhury and S. Sachdev, Phys. Rev. B {\bf 90}, 245136 (2014).
\bibitem{atkinson} W. Atkinsin, A. Kampf and S. Bulut, New J. Phys. {\bf 17} 013025 (2015).
\bibitem{tigran} T. Sedrakyan and A. Chubukov, Phys. Rev. B {\bf 81}, 174536 (2010).
\bibitem{comm} Note that the correlation length in \cite{mike} is defined in their Eq. (4) with extra $1/\sqrt{2}$ compated to our definition, 
  i.e. their $\xi_{\rm AF}=2a$ corresponds to $\xi=2\sqrt{2}a$ in our notations. 
\end{thebibliography}
\end{document}